\documentstyle[epsf,wrapft]{ptptex}

\title{
Formation of Nuclear ``Pasta'' in Cold Neutron Star Matter
}

\author{
Kei {\sc Iida},$^{1,2,}$\footnote{Present address: RIKEN, 2-1 Hirosawa,
Wako, Saitama 351-0198, Japan.}
Gentaro {\sc Watanabe}$^{1}$
and Katsuhiko {\sc Sato}$^{1,3}$
}

\inst{
$^1$Department of Physics, University of Tokyo, 
            7-3-1 Hongo, Bunkyo, Tokyo 113-0033, Japan
\\
$^2$Department of Physics, University of Illinois
            at Urbana-Champaign, 1110 West Green Street,
            Urbana, IL 61801-3080, USA
\\
$^3$Research Center for the Early Universe, University of Tokyo, 
            7-3-1 Hongo, Bunkyo, Tokyo 113-0033, Japan}

\recdate{
\today
}

\abst{
Formation and decay of rod-like and slab-like nuclei, as may be encountered
in the outer parts of neutron stars, are examined within a zero-temperature
stability analysis with respect to perturbations inducing fission and proton 
clustering.  The nonspherical nuclei are found to be stable against these 
perturbations, implying the possibility that the size of the stellar region 
containing such nuclei exceeds the equilibrium prediction.
}

\begin{document}

\maketitle

    Nuclear matter exhibits the coexistence of a liquid phase with a gas phase
at subnuclear densities, due to the attractive force responsible for nuclear
binding.  This feature has been revealed in the vicinity of the critical point 
of the liquid-gas transition through the detection of multifragmentation 
occurring in heavy-ion collision experiments performed at intermediate 
energies.\cite{HIC}  Such a coexistence is believed to be present in the outer
part of a neutron star.\cite{PR}  Since the gravitational coupling is far 
smaller than the Coulomb coupling, the star's gravitational stability requires
nuclear matter to be neutralized by electrons.  The resulting electrically 
neutral matter is usually referred to as neutron star matter.  At sufficiently
low temperatures, relevant to neutron star interiors, long-range Coulomb 
interactions force the system to separate periodically into liquid and gas 
segments of macroscopic size, adding a crystalline property to the liquid-gas
coexistence.  This liquid-gas mixed phase, having two conserved charges, the 
baryon number and the electric charge, is manifested for a finite range of 
pressures.  The presence of electrons controls the proton fraction in the 
liquid and gas phases: As the electron chemical potential increases with 
increasing pressure, a system in $\beta$ equilibrium has the liquid phase 
neutron-enriched and leaves the gas phase almost free of protons.

    At zero temperature, it is believed that the energetically favourable 
configuration of the mixed phase possesses interesting spatial 
structures;\cite{RPW,HSY} the liquid (gas) phase is divided into periodically 
arranged parts of roughly spherical, rod-like or slab-like shape, embedded in 
the gas (liquid) phase and in a roughly uniform electron gas.  Hereafter, we 
refer to these parts composed of the liquid and gas phases as `nuclei' and 
`bubbles', respectively.  Recent calculations, performed within the 
Wigner-Seitz approximation using specific nuclear models,\cite{LRP,O,SOT} 
indicate that at a density of about $10^{14}$ g cm$^{-3}$, rather small 
compared with the saturation density of symmetric nuclear matter, $\rho_{s}
\simeq2.7\times10^{14}$ g cm$^{-3}$, the spatial structure changes from a bcc 
Coulomb lattice of roughly spherical nuclei to a two-dimensional triangular 
lattice of rod-like nuclei.  With the density increased further, it is 
transformed into a layered structure composed of slab-like nuclei and bubbles.
Next, a two-dimensional triangular lattice of rod-like bubbles and a bcc 
Coulomb lattice of roughly spherical bubbles appear in turn.  Finally, at a 
density of about $\rho_{s}/2$, the system dissolves into uniform nuclear 
matter.  These changes of the spatial structure, accompanied by a reduction 
of the total surface area, are governed by the competition between the 
electrostatic and surface energies.  Since slabs and rods look something like
lasagna and spaghetti, the phases with positional order of one and two 
dimensions are often referred to as nuclear ``pasta''.

    In a separate paper,\cite{WIS} we more extensively calculated the 
corresponding equilibrium phase diagrams for zero-temperature neutron star 
matter at subnuclear densities.  In this calculation, we used a compressible 
liquid-drop model developed by Baym, Bethe and Pethick\cite{BBP} (hereafter 
denoted BBP), into which we incorporated uncertainties in the surface 
tension $E_{\rm surf}$ and in the proton chemical potential $\mu_{p}^{(0)}$ in
pure neutron matter.  With an increase of $E_{\rm surf}$, accompanied by an 
increase in the sum of the electrostatic and surface energies, the density 
$\rho_{m}$ at which the system becomes uniform decreases.  There is also a 
tendency for $\rho_{m}$ to decrease with decreasing $\mu_{p}^{(0)}$.  This is
because $-\mu_{p}^{(0)}$ represents the degree to which the gas phase favours 
the presence of protons in itself.  It was shown that while the phases with 
rod-like bubbles and with spherical bubbles can occur only for unrealistically
small $E_{\rm surf}$, the phases with rod-like nuclei and with slab-like 
nuclei survive almost independently of $E_{\rm surf}$ and $\mu_{p}^{(0)}$.

    For these two nuclear phases, there is a direction in which the system is 
translationally invariant.  As noted by Pethick and Potekhin,\cite{PP} this 
situation is geometrically similar to a liquid crystal rather than to a rigid
solid. The elastic properties of the nuclear rods and slabs can thus be 
described by elastic constants used for the corresponding liquid-crystal 
phases, i.e., columnar phases and smectic A phases, respectively.  Pethick and
Potekhin expressed these constants in terms of the electrostatic and surface 
energies.

    We predicted in Ref.\ \citen{WIS} that the phases with rod-like nuclei and
with slab-like nuclei are energetically favoured in the density regime just 
below $\rho_{s}$ and at zero temperature, and it is thus important to 
consider how these nuclei are formed.  Such formation requires simultaneous
migration of an infinite number of nucleons, in contrast to the case of 
ordinary chemical reactions.  This prevents the nuclear system from crossing 
the energy barrier formed between the initial and final states in the 
configuration space via quantum tunnelling.  Instead, ``pasta'' formation can
be driven by instabilities with respect to fluctuations around the initial 
state.

    In this paper, we examine the kinds of instabilities that are involved in 
the formation and decay of rod-like and slab-like nuclei at zero temperature.
As such, we first note an instability with respect to quadrupolar deformation 
of spherical nuclei, as originally investigated in the context of nuclear 
fission by Bohr and Wheeler.\cite{BW}  Pethick and Ravenhall\cite{PR} 
predicted that this instability occurs in the bcc Coulomb lattice and creates
elongated nuclear rods when the volume fraction of the liquid phase reaches 
about $1/8$ in the course of compression.  We also consider an instability 
with respect to proton clustering in uniform matter, as considered by
BBP.\cite{BBP}  This clustering is induced by the isospin symmetry energy.  
At an instability point that the system reaches during decompression, the gain
due to this energy compensates for the gradient and Coulomb energies produced 
by the resulting inhomogeneities, and hence a phase with nuclei of some form
appears.  Pethick and Ravenhall\cite{PR} showed that the critical densities for
instabilities with respect to quadrupolar deformation and proton clustering 
are close to the corresponding equilibrium transition points.

    We extend these considerations to other changes in nuclear shapes.  The 
possible instability with respect to proton clustering in planar and 
cylindrical nuclei tends to divide each nucleus into nuclei of lower dimension,
while the possible fission-like instability of slab-like and rod-like nuclei
tends to lead to the formation of uniform matter and slab-like nuclei, 
respectively.  Using a typical nuclear model, we find that planar and 
cylindrical nuclei are stable with respect to deformation-induced fission and 
proton clustering.  This, together with the finding that these nuclei are 
thermodynamically stable and do not exhibit proton drip, suggests the
possibility that they persist beyond the equilibrium transition points, e.g., 
up (down) to a critical point at which the stable uniform (bcc) phase 
nucleates via quantum tunnelling of the energy barrier in the configuration 
space.  (Note that a critical droplet of the stable phase considered here, 
which, after forming, develops into bulk material, has a finite size, in 
contrast to the critical droplets of the ``lasagna'' and ``spaghetti'' phases.)
Finally, the implications of such persistency for neutron star structure are 
discussed.

    We begin by recalling the mechanism of nuclear fission investigated by Bohr
and Wheeler.\cite{BW}  They regarded a nucleus as a spherical liquid drop of
radius $R$ and total charge $Ze$ in which neutrons and protons are distributed
uniformly, and then calculated the change in the sum of the surface and 
electrostatic energies induced by various kinds of deformations.  Since we are
interested in the limiting case in which the fission barrier vanishes, it is 
sufficient to consider a slight quadrupolar deformation of a spherical drop,
characterized by the distance from the center of the drop to an arbitrary 
point on the surface with polar angle $\theta$: 
\begin{equation} 
X(\theta)=R[1+\alpha_{0}+\alpha_{2}P_{2}(\cos\theta)+\cdots]\ .
 \label{qd}
\end{equation} 
Here, $\alpha_{0}$ is the fractional change in the mean radius of the surface, 
and $\alpha_{2}P_{2}(\cos\theta)$ represents the degree of the 
quadrupolar deformation.  Nuclear saturation leads us to assume the volume of
the drop to be invariant under the deformation (\ref{qd}); we may thus write
$\alpha_{0}=-(1/5)\alpha_{2}^{2}$.  From the condition that the resulting 
change in the sum of the surface and electrostatic energies is zero, we can 
derive the well-known fission-instability relation
\begin{equation} 
E_{C}^{(0)}=2E_{s}^{(0)}\ .
   \label{fission}
\end{equation} 
Here $E_{C}^{(0)}=(3/5)Z^{2}e^{2}/R$ is the Coulomb self energy, and 
$E_{s}^{(0)}=4\pi E_{\rm surf} R^{2}$ is the surface energy with the surface 
tension $E_{\rm surf}$.

    In neutron star matter at zero temperature, these drops form a bcc lattice
embedded in a roughly uniform, neutralizing background of electrons and, at 
densities above neutron drip, in a sea of neutrons.  We assume that the neutron
and proton number densities $n_n$ and $n_p$ are flat outside and inside the 
drop, whereas the electron number density $n_{e}$ is everywhere constant.  (We
ignore here the thickness of the surface layer of the drop.)  In the 
Wigner-Seitz approximation, the electrostatic energy of the Wigner-Seitz cell 
of radius $R_{c}$ reads\cite{PR}
\begin{equation}
E_{C}=E_{C}^{(0)}\left(1-\frac{3}{2}u^{1/3}+\frac{1}{2}u\right)\ ,
 \label{ec}
\end{equation}
where $u=(R/R_{c})^{3}$ is the volume fraction of the drop.  Equilibrium with
respect to $R$ with fixed number densities $n_{i}$ $(i=n,p,e)$ leads to 
\begin{equation}
2E_{C}=E_{s}^{(0)}\ .
 \label{sizeeq}
\end{equation}

    By using conditions (\ref{fission}) and (\ref{sizeeq}) up to $O(u^{1/3})$,
we obtain the fission-instability criterion, $u=1/8$, derived by Pethick and 
Ravenhall.\cite{PR}  Here we have noted that corrections to condition 
(\ref{fission}) due to the lattice are of order $u$.\cite{Brandt}  Since $u$
generally increases with increasing density, we can uniquely determine the
density at which the drop becomes unstable with respect to the deformation 
(\ref{qd}).

    We next derive a fission-instability condition appropriate for a nuclear
rod, composed of uniformly distributed neutrons and protons (proton charge 
density $\rho$) as well as having a circular section (sectional radius $R$) 
and an infinitely long axis.  For this rod, we consider a small sectional 
deformation of the quadrupole type that is uniform in the direction of the
axis.  This deformation is characterized by the distance, measured on a 
horizontal section, from the axis to a point located on the surface with a 
given angle $\theta$:
\begin{equation}
X(\theta)=R(1+\alpha_{0}+\alpha_{2}\cos2\theta+\cdots)\ .
 \label{rd}
\end{equation}
Here, $\alpha_{0}$ is the fractional change in the mean sectional radius of 
the rod, and $\alpha_{2}\cos2\theta$ represents the degree of the 
quadrupole-type deformation.  Subject to the condition of constant sectional 
area, we obtain $\alpha_{0}=-(1/4)\alpha_{2}^{2}$.

    Proceeding with the argument for the rod in a vacuum, we are inevitably 
led to an infinite value of the electrostatic energy per unit length.  For 
this reason, we include the effect of a triangular lattice formed by the rods 
immersed in a uniform neutralizing background of electrons and in a sea of 
neutrons.  We here again utilize the Wigner-Seitz approximation.  The 
electrostatic energy for a Wigner-Seitz cell of sectional radius $R_{c}$ can
then be expressed per unit length as 
\begin{equation}
E_{C}=\frac{(\pi\rho R^{2})^{2}}{2}(-\ln u-1+u)\ ,
 \label{ecrod}
\end{equation}
where $u=(R/R_{c})^{2}$ is the volume fraction of the rod.  The deformation 
(\ref{rd}) produces a change in the electrostatic energy from this $E_{C}$ 
given by
\begin{equation}
\delta E_{C}=(\pi\rho R^{2})^{2}\left(-\frac{1}{2}+u\right)
\alpha_{2}^{2}\ .
 \label{decrod}
\end{equation}
Next, we can obtain the surface energy per unit length of the undeformed rod 
and its change due to the deformation (\ref{rd}) as 
\begin{equation}
E_{s}^{(0)}=2\pi E_{\rm surf} R\ ,
 \label{esrod}
\end{equation}
\begin{equation}
\delta E_{s}=\frac{3}{4}E_{s}^{(0)}\alpha_{2}^{2}\ .
 \label{desrod}
\end{equation}

    By combining the instability condition $\delta E_{C}+\delta E_{s}=0$, 
determined by Eqs.\ (\ref{decrod}) and (\ref{desrod}), with the 
size-equilibrium condition (\ref{sizeeq}), in which $E_{C}$ and $E_{s}^{(0)}$
are now given by Eqs.\ (\ref{ecrod}) and (\ref{esrod}), we conclude that the
rod is stable with respect to the deformation (\ref{rd}) for $0 \leq u \leq 1$.
Recall that this deformation is uniform in the direction of the axis.  Any 
small deviation from this uniformity with a fixed nucleon density and volume of
the rod, however, results in further energy loss.  This is because an increase
in the surface energy, which is proportional to the deviation, cannot be 
compensated for by a change in the Coulomb energy, which is of second order in
the deviation.  We remark in passing that deformations of the liquid-crystal
type as considered by Pethick and Potekhin\cite{PP} are elastic, and are not 
accompanied by the sectional distortion considered here.

    In the case of a nuclear slab of thickness $2R$ contained in a cell of
thickness $2R_{c}$, there is no quadrupole-type deformation corresponding to
(\ref{qd}) and (\ref{rd}).  All the deformations that hold fixed the nucleon
density and volume of the slab have already been examined in the context of 
elastic deformations by Pethick and Potekhin;\cite{PP} the slab experiences
a restoring force for these deformations.

    We next turn to the condition for instability with respect to proton 
clustering in uniform nuclear matter neutralized and $\beta$ equilibrated by 
electrons.  This condition was obtained by BBP\cite{BBP} by expanding the 
energy density functional $E[n_{i}({\bf r})]$ $(i=n,p,e)$ of the system with 
respect to small density fluctuations $\delta n_{i}({\bf r})$ around the 
homogeneous state.  The contribution of first order in $\delta n_{i}$ vanishes
due to the equilibrium of the unperturbed homogeneous system, while the second
order contribution can be described in the spirit of the extended Thomas-Fermi
model for finite nuclei as\cite{BBP}
\begin{equation}
E-E_{0}=\frac{1}{2}\int\frac{d{\bf q}}{(2\pi)^{3}}v(q)
|\delta n_{p}({\bf q})|^{2}\ ,
 \label{TF}
\end{equation}
where $E_{0}$ is the ground-state energy, $\delta n_{p}({\bf q})$ is the 
Fourier transform of $\delta n_{p}(\bf r)$, and $v(q)$ is the potential of 
the effective interaction between protons, as given by
\begin{equation}
v(q)=v_{0}+\beta q^{2}+\frac{4\pi e^{2}}{q^{2}+k_{\rm TF}^{2}}\ .
 \label{v}
\end{equation}
Here, 
\begin{equation}
v_{0}=\frac{\partial\mu_{p}}{\partial n_{p}}-
\frac{(\partial\mu_{p}/\partial n_{n})^{2}}
     {\partial\mu_{n}/\partial n_{n}}\ ,
 \label{v0}
\end{equation}
\begin{equation}
\beta=\frac{2}{n_{\rm NM}}(B_{pp}+2B_{np}\zeta+B_{nn}\zeta^{2})\ ,
 \label{beta}
\end{equation}
\begin{equation}
\zeta=-\frac{\partial \mu_{p}/\partial n_{n}}
            {\partial \mu_{n}/\partial n_{n}}\ ,   
 \label{zeta}
\end{equation}
with $\mu_{n(p)}$ the neutron (proton) chemical potential, $n_{\rm NM}$ the 
saturation density of symmetric nuclear matter, and $B_{ij}$ the matrix 
determining the gradient term in $E$, and $k_{\rm TF}\approx 0.3 n_{e}^{1/3}$ 
is the Thomas-Fermi screening length of the ultrarelativistic electrons,
modifying the Coulomb term in $E$.  The potential $v(q)$ takes a minimum value
$v_{\rm min}$ at $q=Q$, where 
\begin{equation}
Q^{2}=\left(\frac{4\pi e^{2}}{\beta}\right)^{1/2}-k_{\rm TF}^{2}\ ,
 \label{Q}
\end{equation}
\begin{equation}
v_{\rm min}=v_{0}+2(4\pi e^{2}\beta)^{1/2}-\beta k_{\rm TF}^{2}\ .
 \label{vmin}
\end{equation}

    In the energy expansion up to second order in $\delta n_{i}$, the condition
that uniform nuclear matter becomes unstable with respect to proton clustering
reads $v_{\rm min}=0$.\cite{BBP}  Generally, $v_{\rm min}$ is controlled by 
the bulk contribution $v_{0}$, which increases as the proton fraction (or, 
equivalently, the density) increases, while the gradient and 
Coulomb terms, which tend to suppress the instability, are small 
positive-definite corrections to $v_{0}$.\cite{PRL}  Such a density dependence
of $v_{0}$ ensures the presence of a density above (below) which the matter is
stable (unstable) with respect to proton clustering.

    For possible proton clustering in phases with nuclear rods and nuclear 
slabs, we may assume that the density fluctuations $\delta n_{i}$ leading to 
proton clustering are confined within the nuclei, as it requires too much 
energy to produce proton clustering in the surrounding neutron gas.  Thus, it 
is useful to consider the effective interaction between protons in the nuclei,
corresponding to Eq.\ (\ref{v}) in uniform nuclear matter.  The bulk term 
$v_{0}$ can now be calculated by substituting the neutron and proton number 
densities inside the nuclei into Eq.\ (\ref{v0}).  We use the condition 
$v_{0}=0$ when estimating the density at which nuclear matter in the slab or 
rod becomes unstable with respect to proton clustering.  We remark that
rigorous extension of the proton effective interaction in uniform nuclear 
matter to that in the nuclei requires equilibrium neutron and proton 
distributions, which are inhomogeneous throughout the regions containing
protons (see, e.g., Ref.\ \citen{O}), and inclusion of the gradient and 
Coulomb correction terms.  This Coulomb term depends explicitly on the spatial
scales $R$ and $R_{c}$.

    We now proceed to calculate the densities at which instabilities arise with
respect to deformation-induced fission of spherical nuclei and proton 
clustering in both uniform matter and nonspherical nuclei.  For this purpose, 
we use the compressible liquid-drop model for nuclei that we constructed in 
Ref.\ \citen{WIS}.  To obtain this model, we modified the BBP model by using 
$\mu_{p}^{(0)}=-C_{1} n_{n}^{2/3}$ [Eq.\ (4) in Ref.\ \citen{WIS}] and 
$E_{\rm surf}=C_{2} \tanh(3.5~{\rm MeV}/\mu_{n}^{(0)}) E_{\rm surf}^{\rm BBP}$
[Eq.\ (6) in Ref.\ \citen{WIS}], where $\mu_{n}^{(0)}$ is the neutron chemical
potential in the neutron gas, and $E_{\rm surf}^{\rm BBP}$ is the BBP-type 
surface tension [Eq.\ (7) in Ref.\ \citen{WIS}].  Here, we set the parameters 
$C_{1}$ and $C_{2}$ as $C_{1}=400$ MeV fm$^{2}$ and $C_{2}=1$, corresponding to
typical values of $\mu_{p}^{(0)}$ and $E_{\rm surf}$ used in recent studies
(see Figs.\ 1 and 2 in Ref.\ \citen{WIS}).  This model, allowing for a bcc 
lattice of spherical nuclei or bubbles, a two-dimensional triangular lattice 
of cylindrical nuclei or bubbles, and a layered lattice of planar nuclei, 
exhibits the following first order phase transitions in the ground-state 
neutron star matter: spherical nuclei $\to$ cylindrical nuclei $\to$ planar 
nuclei $\to$ uniform matter (with increasing density).  The associated 
discontinuities in the volume fraction $u$ and proton fraction $x$ of the 
nuclear matter region can be seen from Fig.\ 1, displaying $u$ and $x$ as 
functions of the baryon density $n_{b}$.  The important point is that the 
transition density, $n_{b}\approx0.079$ fm$^{-3}$, from the phase with 
spherical nuclei to that with cylindrical nuclei is close to the fission-like
instability point $u\approx1/8$ or, equivalently, $n_{b}\approx0.061$ 
fm$^{-3}$.  This result is consistent with the analysis of Pethick and 
Ravenhall.\cite{PR}

\begin{figure}[htb]
   \parbox{\halftext}{
                \vspace{-1.52cm}
                \epsfxsize=7.2cm 
                \epsfysize=6.4cm 
                \centerline{\epsfbox{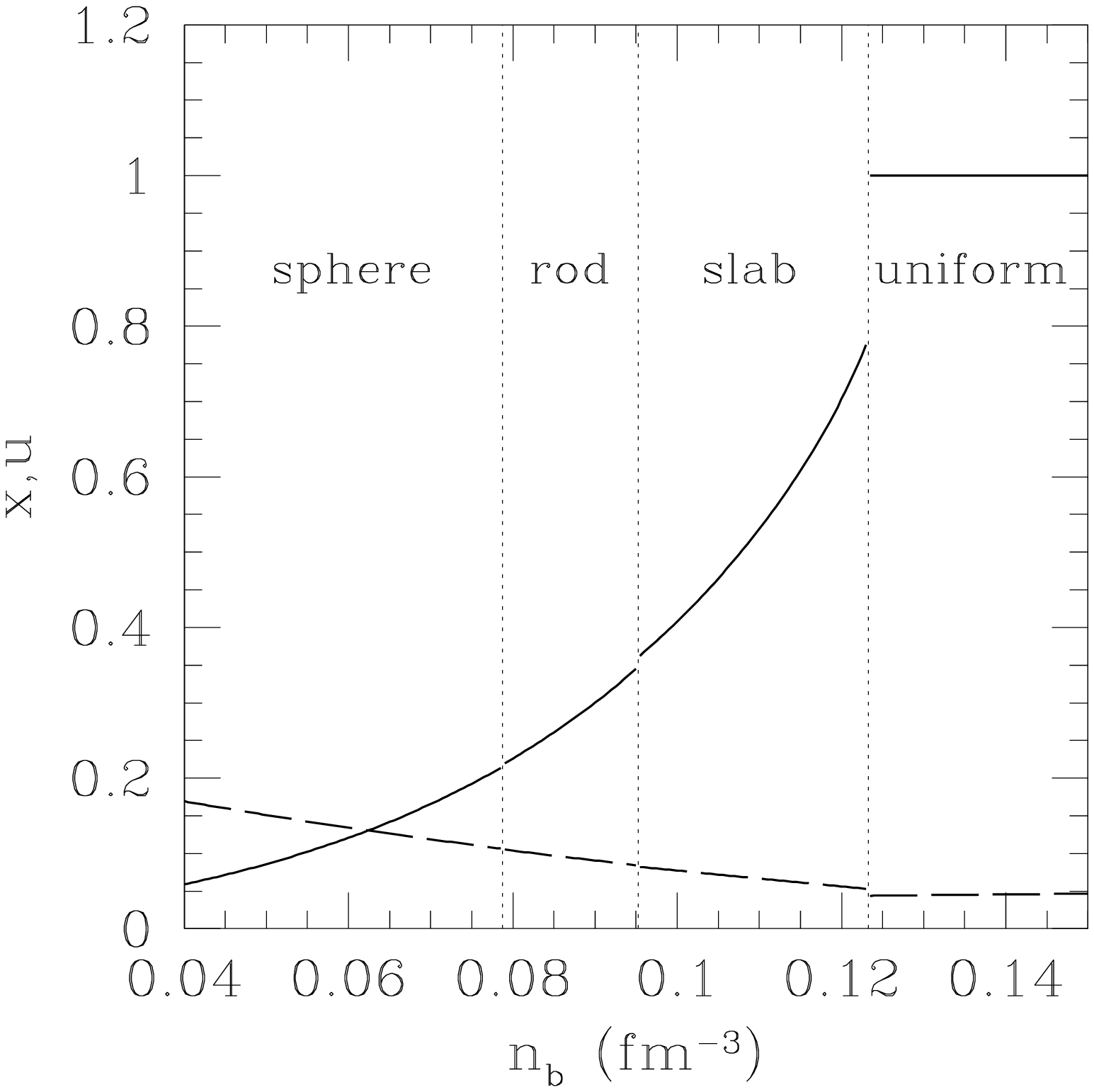}}
\caption{
The volume fraction $u$ and the proton fraction $x$ of the nuclear 
matter region, calculated for the ground-state neutron star matter.  The solid
curve represents the volume fraction, and the dashed curve represents the 
proton fraction.
}}
\hspace{4mm}
   \parbox{\halftext}{
                \vspace{-0.7cm}
                \epsfxsize=7.2cm 
                \epsfysize=6.4cm 
                \centerline{\epsfbox{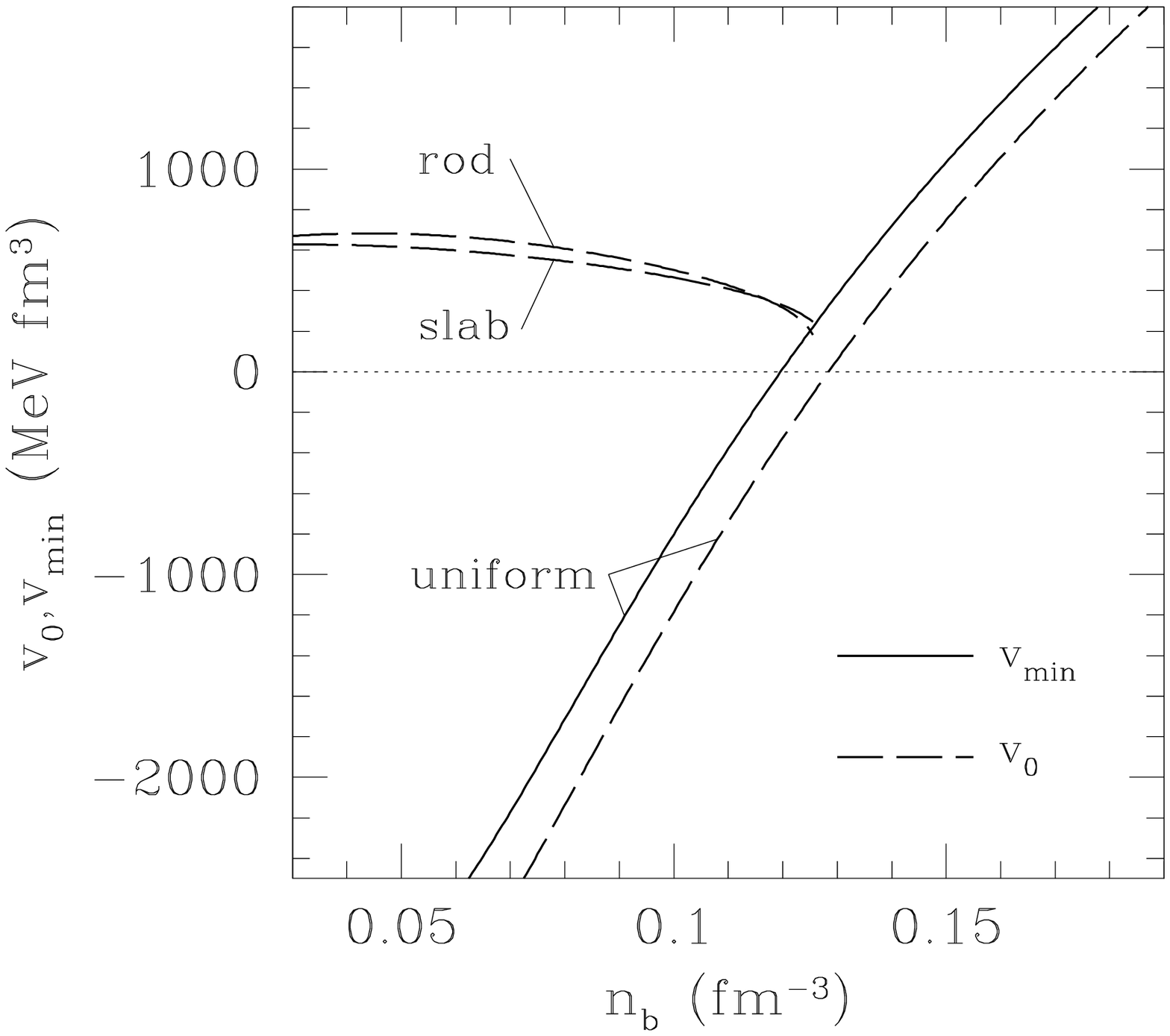}}
\caption{
The minimum value $v_{\rm min}$ and the bulk contribution $v_{0}$ of 
the effective potential between protons lying in the nuclear matter region, 
calculated for the phases with cylindrical nuclei and planar nuclei, as well as
for the uniform phase.  The solid curve represents the result for 
$v_{\rm min}$, and the dashed curves represent the results for $v_{0}$.
}}
\end{figure}

    Figure 2 shows the results for $v_{\rm min}$, Eq.\ (\ref{vmin}), and 
$v_{0}$, Eq.\ (\ref{v0}), calculated from the compressible liquid-drop model
used here.  In this calculation, we have set $n_{\rm NM}=0.17$ fm$^{-3}$ and 
$B_{nn}=B_{np}=B_{pp}=8.05$ MeV fm$^{2}$ so as to reproduce the gradient term 
in model I of Oyamatsu;\cite{O} this term is consistent with the experimental
masses and radii of normal nuclei.  From Fig.\ 2 we can estimate the critical 
density at which proton clustering occurs in uniform matter to be $\approx
0.120$ fm$^{-3}$.  This value agrees well with the phase-equilibrium point, 
$n_{b}\approx0.123$ fm$^{-3}$, which can be obtained from Fig.\ 1, a feature 
consistent with the result of Pethick and Ravenhall.\cite{PR}  The critical 
wavelength $2\pi Q^{-1}$ is obtained as $2\pi Q^{-1}\sim$ 20 fm.  This is 
comparable to an internuclear spacing of $\sim 2R_{c}$; the cell size $R_{c}$,
calculated for the ground state matter with the present nuclear model, is 
shown in Fig.\ 3.  This suggests the eventual formation of a lattice of nuclei
of some form.

    We can observe from Fig.\ 2 that for cylindrical and planar nuclei, the
bulk term $v_{0}$ increases with decreasing density and hence does not become
negative, in contrast to the case of uniform nuclear matter.  This implies 
that nonspherical nuclei are stable with respect to proton clustering.  The 
observed density dependence of $v_{0}$ results from 
\begin{wrapfigure}{r}{6.6cm}   
                \vspace{-0.8cm}
                \epsfxsize=7.0cm 
                \epsfysize=6.2cm 
                \centerline{\epsfbox{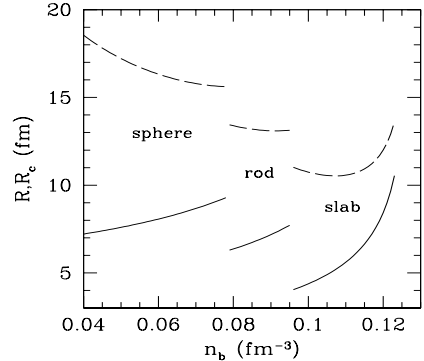}}
\caption{
Size of a nucleus, $R$, and of a Wigner-Seitz cell, $R_{c}$, 
calculated for the ground-state neutron star matter.  The solid lines are the
results for $R$, and the dashed lines are the results for $R_{c}$.
}
\end{wrapfigure}
the facts that $v_{0}$ 
increases with increasing proton fraction $x$ and that, as seen in Fig.\ 1, 
$x$ increases with decreasing density.  Moreover, calculations of the 
difference between the proton chemical potentials in liquid (nuclear matter) 
and gas (neutron matter) regions and of the compressibility of the system 
employing the present nuclear model and following a line of argument used by 
BBP\cite{BBP} show that there is no indication of the drip of protons out of 
nonspherical nuclei nor a thermodynamic instability of the system including 
nonspherical nuclei.  Consequently, we can conclude that within the present 
nuclear model, no channel leading to the decay of nonspherical nuclei is 
opened by the possible instabilities analyzed here.

    Finally, we consider a situation in which phases with nuclear rods and 
nuclear slabs are formed in a neutron star.  Under the subsequent spin-down 
of the star, matter including nonspherical nuclei would be compressed in the 
equatorial region and decompressed in the polar region.  Additionally, 
accretion of material from a companion star, if occurring, would act to 
compress the matter in the entire region of the star.  During decompression, 
nonspherical nuclei might not separate into finite nuclei until a critical 
metastability with respect to quantum-tunnelling nucleation of the usual bcc 
phase with roughly spherical nuclei is realized, while during compression, they
might persist up to a critical density at which uniform matter nucleates 
quantum mechanically in a metastable system or the volume fraction $u$ 
becomes sufficiently close to unity for the system to melt into uniform
matter.  The possible presence of such critical metastability would act to 
enlarge the stellar region containing the cylindrical and planar nuclei, as 
compared with that predicted using the equilibrium configuration.  In order to
estimate the size of this region, we should take into account dynamical 
aspects of nucleation processes and finite-temperature effects; the latter 
tend to reduce the critical metastability by promoting the quantum decay of 
nonspherical nuclei, and, as we demonstrated in Ref.\ \citen{WIS}, by melting
the low-dimensional lattice of nonspherical nuclei via elastic deformations of
long wavelengths.

\section*{Acknowledgements}

    We are grateful to Professor Takeo Izuyama for helpful discussions.  KI 
would like to thank the Department of Physics of the University of Illinois at
Urbana-Champaign for hospitality during the preparation of this manuscript. 
This work was supported in part by Grants-in-Aid for Scientific Research 
provided by the Ministry of Education, Science and Culture of Japan through 
Research Grant Nos.\ 07CE2002 and 10-03687, and in part by the National Science
Foundation Grant No.\ PHY98-00978.

\end{document}